\documentclass{amsart}
\usepackage{amssymb,amsmath,amscd, amsthm,stmaryrd,verbatim, mathrsfs}

\newtheorem{Thm}{Main Theorem}
\newtheorem{theorem}{Theorem}[section]

\newtheorem{Lem}[theorem]{Lemma}
\newtheorem{Claim}[theorem]{Claim}
\newtheorem{Prop}[theorem]{Proposition}

\theoremstyle{definition}
\newtheorem{Def}[theorem]{Definition}

\theoremstyle{remark}
\newtheorem{rmk}[theorem]{Remark}

\numberwithin{equation}{section}

\newcommand{\bb}{\mathbb}
\newcommand{\ms}{\mathscr}
\newcommand{\mr}{\mathrm}
\newcommand{\frk}{\mathfrak}

\usepackage{hyperref}
\begin{document}

\title{The $\mr{Sp}(1)$-Kepler Problems}
\author{Guowu Meng}

\address{Department of Mathematics, Hong Kong Univ. of Sci. and
Tech., Clear Water Bay, Kowloon, Hong Kong}
%\curraddr{Department of Mathematics and Statistics}

\email{mameng@ust.hk}

%    General info

\date{May 7, 2008}



\begin{abstract} Let $n\ge 2$ be a positive integer.
To each irreducible representation $\sigma$ of $\mathrm{Sp}(1)$, an
$\mathrm{Sp}(1)$-Kepler problem in dimension $(4n-3)$ is constructed and
analyzed. This system is super integrable and when $n=2$ it is
equivalent to a generalized MICZ-Kepler problem in dimension five.
The dynamical symmetry group of this system is $\widetilde {\mathrm
O}^*(4n)$ with the Hilbert space of bound states ${\mathscr H}(\sigma)$
being the unitary highest weight representation of $\widetilde {\mathrm
{O}^*}(4n)$ with highest weight
$$(\underbrace{-1, \cdots,  -1}_{2n-1},  -(1+\bar\sigma)),$$
which occurs at the right-most nontrivial reduction point in the
Enright-Howe-Wallach classification diagram for the unitary highest
weight modules. Here $\bar\sigma$ is the highest weight of $\sigma$.
Furthermore, it is shown that the correspondence
$\sigma\leftrightarrow \mathscr H(\sigma)$ is the theta-correspondence
for dual pair $(\mathrm{Sp}(1), \mathrm{O}^*(4n))\subseteq\mathrm{Sp}_{8n}(\mathbb R)$.
\end{abstract}

\maketitle

\section {Introduction}
The Kepler problem is a well-known physics problem in dimension
three about two bodies which attract each other by a force
proportional to the inverse square of their distance. What is less
known about the Kepler problem is the fact that it is super
integrable\footnote{A physics model is called {\em super integrable}
if the number of independent symmetry generators is bigger than the
number of degree of freedom. For the Kepler problem, the degree of
freedom is $3$ and the number of independent symmetry generators is
$5$.} at both the classical and the quantum level, and belongs to a
big family of super integrable models. One interesting such family
is the family of \emph{MICZ-Kepler problems} \cite{MC70, Z68}.

In our recent research on the Kepler problem \cite{meng05,MZ07,M07,
meng08a, meng08b}, a dominant theme is the construction of new super
integrable models of Kepler type on the one hand and the exhibition
of the close relationship of these super integrable models with
certain unitary highest weight modules for a real non-compact Lie
group of Hermitian type on the other hand. Depending on the
interests of the readers, one may view this investigation either as
a journey to discover new super integrable models or as an effort to
better understand the geometry of certain unitary highest weight
modules. Here we continue this theme.

In Ref. \cite{meng08a}, a new family of super integrable models of
the Kepler type
--- {\em the $\mr O(1)$-Kepler problems} --- has been constructed and
analyzed, and their intimate relationship with theta-correspondence
has been uncovered. Their complex analogue --- {\em the $\mr
U(1)$-Kepler problems} --- have been constructed and analyzed in
Ref. \cite{meng08b}. The purpose here is to construct and analyze
their quaternionic analogues
--- {\em the $\mr{Sp}(1)$-Kepler problems}.

Recall that, in the construction of the $\mr O(1)$-Kepler problems
in dimension $n$, the canonical bundle
$$
\mr{O}(1)\to \mr{S}^{n-1}\to\bb R\mr{P}^{n-1}
$$
plays a pivotal role. Here the corresponding bundle is
$$
\mr{Sp}(1)\to \mr{S}^{4n-1}\to\bb H\mr{P}^{n-1}. $$ Later we shall
demonstrate that, a model with $n=2$ constructed in this paper is
equivalent to an Iwai's $\mr{SU}(2)$-Kepler problem \cite{Iwai90}.
That is why the models constructed here are called the \emph{the
$\mr{Sp}(1)$-Kepler problems}.

\subsection{Statement of the Main Results}
Before stating our main result, let us fix some notations:
\begin{itemize}
\item $n$ --- an integer which is at least 2;
\item $\sigma$ --- an irreducible representation of $\mr{Sp}(1)$,
it also denotes the underlying representation space of $\sigma$;
\item $\bar \sigma$ --- the highest weight of $\sigma$, so it is a non-negative integer;
\item $\tilde {\mr U}(2n)$ --- the nontrivial double cover of $\mr
U(2n)$;
\item $\widetilde{\mr {O}^*}(4n)$ --- the nontrivial double cover of ${\mr {O}^*}(4n)$;
\item $\kappa$ --- a half integer;
\item $l$ --- a non-negative integer;
\item $\mathcal R_l^\kappa(\sigma)$ ---  the highest weight module of
$\tilde{\mr U}(2n)$ with highest weight
$$(l+\bar \sigma+\kappa, l+\kappa, \underbrace{\kappa, \cdots,  \kappa}_{2n-2});$$
\item $\overline{\mathcal R_l^\kappa(\sigma)}$ ---  the highest weight module of
$\tilde{\mr U}(2n)$ with highest weight
$$(\underbrace{-\kappa, \cdots,  -\kappa}_{2n-2}, -(l+\kappa), -(l+\bar \sigma+\kappa)).$$

\end{itemize}

We are now ready to state the main  results on the
$\mr{Sp}(1)$-Kepler problems.
\begin{Thm}\label{T:main} Let $n\ge 2$ be an integer, $\sigma$ an irreducible representation of $\mr{Sp}(1)$, and
$\bar \sigma$ the highest weight of $\sigma$. For the
$(4n-3)$-dimensional ${\mr {Sp}}(1)$-Kepler problem with magnetic
charge $\sigma$, the following statements are true:

1) The bound state energy spectrum is
$$
E_I=-{1/2\over (I+n+{\bar\sigma\over 2})^2}
$$ where $I=0$, $1$, $2$, \ldots

2) There is a natural unitary action of $\widetilde{\mr {O}^*}(4n)$
on the Hilbert space ${\ms H}(\sigma)$ of bound states, which
extends the manifest unitary action of $\mr {Sp}(n)$. In fact, ${\ms
H}(\sigma)$ is the unitary highest weight representation of
$\widetilde {\mr {O}^*}(4n)$ with highest weight
$$(\underbrace{-1, \cdots,  -1}_{2n-1},  -(1+\bar\sigma)).$$

3) When restricted to the maximal compact subgroup $\tilde {\mr
U}(2n)$, the above action yields the following orthogonal
decomposition of $\ms H(\sigma)$:
$$
{\ms H}(\sigma)=\hat\bigoplus _{I=0}^\infty\,{\ms H}_I(\sigma)
$$ where, as the irreducible $\tilde {\mr {U}}(2n)$-representation,
${\ms H}_I(\sigma)$ is the unitary highest weight representation
with highest weight
$$(\underbrace{-1, \cdots,  -1}_{2n-2}, -(1+I), -(1+I+\bar\sigma)).$$

4) ${\ms H}_I(\sigma)$ in part 3) is the energy eigenspace with
eigenvalue $E_I$ in part 1).

5) The correspondence between $\sigma$ and $\ms H(\sigma)$ is the
theta-correspondence\footnote{See Ref. \cite{Howe} for details on
reductive dual pairs and theta-correspondence.} for dual pair
$(\mr{Sp}(1), \mr{O}^*(4n))\subseteq \mr{Sp}_{8n}(\bb R)$.
\end{Thm}
For readers who are familiar with the Enright-Howe-Wallach
classification diagram \cite{EHW82} for the unitary highest weight
modules, we would like to point out that, the unitary highest weight
module identified in part 2) of this theorem occurs at the rightmost
nontrivial reduction point of the classification diagram. Note that,
part 3) of the above theorem is a \emph{multiplicity free} $K$-type
formula.

In section \ref{S:model}, we introduce a super integrable model for
each integer $n\ge 2$ and each representation of
$\mr{Sp}(1)=\mr{SU}(2)$. Here the scalar potential is (uniquely)
determined by the requirement that the radial Schrodinger equation
can be solved in terms of generalized Laguerre polynomials. For
$n=2$, these models are shown to be equivalent to (the quantum
analogues of) Iwai's $\mr{SU}(2)$-Kepler problems.

In section \ref{S:analysis}, we give a detailed analysis of the
models and finish the proof of main theorem \ref{T:main}. Here, the
bound state problem is solved with the help of the well-known
branching rules for compact symmetric pairs, and then the large
hidden symmetry is exhibited. For the dynamic symmetry, a crucial
connection of these models with harmonic oscillators is obtained and
subsequently exploited. Through these detailed analysis, the models
can be seen clearly to be super integrable and share the
characteristic features of the original Kepler problem; together
with the fact that they are equivalent to Iwai's $\mr{Sp}(1)$-Kepler
problems when $n=2$, we refer these models as $\mr{Sp}(1)$-Kepler
problems.

As expertly pointed out by the referee, the models introduced here
and in Refs. \cite{meng08a, meng08b} can be alternatively obtained
from the harmonic oscillators via the quantum symmetry reduction.
Our method seems to work more generally, that is why we can make
progress in Ref. \cite{meng05}.

It is worth to remark at this point that the super integrable models
constructed in this paper and our other papers is perhaps just a
small part of {\em the universe of super integrable models of Kepler
type}. In fact, a new super integrable model with $E_{7(-25)}$ as
its dynamic symmetry group will appear in a forthcoming joint paper
with J.S. Li.

\section{The models}\label{S:model}
Let $n\ge 2$ be an integer, $\bb H^n_*=\bb H^n\setminus\{0\}$, and
$\sigma$ an irreducible unitary representation of $\mr{Sp}(1)$.
Denote by $\bar \sigma$ be the highest weight of $\sigma$.

Consider the principal bundle
$$
{\mr {Sp}}(1)\to \bb H^n_*\to \widetilde{\bb HP^n}
$$
where $\widetilde{\bb HP^n}\cong \bb R_*\times {\bb H}P^{n-1}$ is
the quotient space of  $\bb H^n_*$ under the equivalence relation
$Z\sim \alpha\cdot Z$, $\alpha\in \mr{Sp}(1)$. We shall always
assume the Euclidean metric on $\bb H^n_*$  and the resulting
quotient Riemannian metric on $\widetilde{\bb HP^n}$.

We use $(\rho, \Phi)$ to denote the polar coordinates on
$\widetilde{\bb HP^n}$ and $\gamma_\sigma$ to denote the vector
bundle associated to representation $\sigma$. Then $\rho([Z])=|Z|$,
$\Phi$ is a coordinate on $\bb HP^{n-1}$, and $\gamma_\sigma$ is a
hermitian line bundle over $\widetilde{\bb HP^n}$ with a natural
hermitian connection $\mathcal A$. Here is a technical lemma which
will be used later.

\begin{Lem}\label{1stLem} 1) Let $Z\in \bb H^n_*$ and $\rho=|Z|=\sqrt{\bar Z\cdot
Z}$. Then we have the following identity for the Euclidean metric on
$\bb H^n_*$:
$$
|dZ|^2=d\rho^2+\rho^2\left(ds^2_{FS}+\left({Im(\bar Z\cdot dZ)\over
|Z|^2}\right)^2\right)
$$ where ``$Im$'' in $Im(\bar Z\cdot dZ)$ stands for ``the imaginary part
of", and $ds^2_{FS}$ is the Fubini-Study metric\footnote{It is the
quotient metric on $\mr{S}^{4n-1}/\mr{Sp}(1)$. On ${\bb HP}^1={\mr
S}^4$, it is $1\over 4$ times the standard round
metric.}\label{foot} on ${\bb H}P^{n-1}$, i.e.,
\begin{eqnarray}
ds^2_{FS}={|dZ|^2\over |Z|^2}-{|\bar Z\cdot dZ|^2\over |Z|^4}.
\end{eqnarray} Consequently, the quotient metric on $\widetilde{\bb
HP^n}$ is
\begin{eqnarray}
ds^2_{\widetilde{\bb HP^n}}=d\rho^2+\rho^2 ds^2_{FS}.
\end{eqnarray}

2) If we take the invariant Riemannian metric on $\mr{Sp}(n)$ as the
restriction of the Euclidean metric on $\bb H^{n^2}$, then the
resulting quotient metric on  $\bb HP^{n-1}={\mr{Sp}(n)\over
\mr{Sp}(n-1)\times\mr{Sp}(1)}$ is twice of the Fubini-Study metric.
Consequently,
\begin{eqnarray}\label{LapFor}\Delta_{\mathcal A}|_{\bb HP^{n-1}}=2\left(c_2[\mr
{Sp}(n)]-c_2[\mr{Sp}(1)]|_\sigma\right)\end{eqnarray} where
$\Delta_{\mathcal A}|_{\bb HP^{n-1}}$ is the (non-negative) Laplace
operator acting on sections of $\gamma_\sigma|_{\bb HP^{n-1}}$,
$c_2[\mr{Sp}(n)]$ is the Casimir operator of $\mr {Sp}(n)$, and
$c_2[\mr {Sp}(1)]|_\sigma$ is the value of the Casimir operator of
$\mr {Sp}(1)$ at $\sigma$.

\end{Lem}
\begin{proof} 1) Let $Z=(Z_1, \cdots. Z_n)\in \bb H^n_*$. The Euclidean metric on
$\bb H^n_*$ is just $|dZ|^2=dZ\cdot d\bar Z$. Since
$\rho^2=|Z|^2=Z\cdot \bar Z$, $2\rho d\rho=Z\cdot d\bar Z+\bar
Z\cdot dZ$, and $$ d\rho^2={2|Z\cdot d\bar Z|^2+(Z\cdot d\bar
Z)^2+(\bar Z\cdot dZ)^2\over 4|Z|^2}.
$$ A computation shows that
$$
|dZ|^2=d\rho^2+\rho^2\left(ds^2_{FS}+\left({Im(\bar Z\cdot dZ)\over
|Z|^2}\right)^2\right).
$$
The right $\mr{Sp}(1)$ action on $\bb H^n_*$ generates a rank $3$
sub-bundle of $T\bb H^n_*$. The fiber of this sub-bundle at point
$Z_0$ is $\mr{span}_{\bb R}\{(Z_0, Z_0i), (Z_0, Z_0j), (Z_0,
Z_0k)\}$. Let $V_{Z_0}$ be the orthogonal complement of
$\mr{span}_{\bb R}\{Z_0i, Z_0j, Z_0k\}$ in ${\bb H}^n$($=\bb
R^{4n}$), so
$$
V_{Z_0}=\left\{W\in \bb H^{n}\mid Im(\bar Z_0\cdot W)=0 \right\}.
$$

To finish the proof of part 1), we just need to check that quadratic
form
$$\left.\left(Im(\bar Z\cdot dZ)\right)^2\right|_{Z_0}$$ vanishes
when restricted to $Z_0\times V_{Z_0}$. Indeed, if $v:=(Z_0, W)\in
\{Z_0\}\times V_{Z_0}$, then
$$
\left.\left(Im(\bar Z\cdot dZ)\right)^2\right|_{Z_0}(v, v)=(Im(\bar
Z_0\cdot W))^2=0.
$$

2) The Euclidean metric on $\bb H^{n^2}$ is $\mr{tr}(dM^\dag\, dM)$,
so the Riemannian metric on $\mr{Sp}(n)$ is
$ds^2_{\mr{Sp}(n)}=\mr{tr}(dg^\dag\, dg)$. Since the action of $\mr
{Sp}(n)$ on $\bb HP^{n-1}={\mr{Sp}(n)\over
\mr{Sp}(n-1)\times\mr{Sp}(1)}$ is transitive and both the
Fubini-Study metric and the quotient metric here are $\mr{Sp}(n)$
invariant, we just need to verify the statement at point $p:=[1,
0,\cdots ,0]\in \bb HP^{n-1}$. Let $I\in \mr{Sp}(1)$ be the identity
matrix.

In the following, if $X$ is a Riemannian manifold, we shall use
$\langle, \rangle|_X$ to denote the Riemannian inner product on the
tangent space of $X$ at some point.

Let $I\in \mr{Sp}(1)$ be the identity matrix. Let $u, v\in
T_I\mr{Sp}(n)$ be two tangent vectors which are orthogonal to
$T_I(\mr{Sp}(n-1)\times\mr{Sp}(1))$. Then, there are $a, b\in \bb
H^{n-1}$ such that
$$
u=\left(I, \left( \begin{matrix}0 & -a^\dag\cr a & 0\end{matrix}
\right)\right), \quad v=\left(I, \left( \begin{matrix}0 & -b^\dag
\cr b & 0\end{matrix} \right)\right),
$$ so $\langle u, v\rangle|_{\mr{U}(n)}=a^\dag b+b^\dag a=2Re(a^\dag b)$.
Let $$\tilde p=(1, \underbrace{0, \cdots, 0}_{n-1})^T\in
\mr{S}^{4n-1},$$ and $\tilde u$, $\tilde v$ be the image of $u$, $v$
respectively under the linearlization at $I$ of the quotient map
\begin{eqnarray}
{\mr{Sp}}(n) &\to & \mr {S}^{4n-1}\cr A&\mapsto & A\tilde
p\;.\nonumber
\end{eqnarray} Then
$$
\tilde u=\left(\tilde p, \left( \begin{matrix}0 \cr a \end{matrix}
\right)\right), \quad \tilde v=\left(\tilde p, \left(
\begin{matrix}0 \cr b \end{matrix} \right)\right),
$$ so $\langle \tilde u, \tilde v\rangle|_{\mr{S}^{4n-1}}=Re(a^\dag b)$.
Therefore,
\begin{eqnarray}\label{2wice}
\langle u, v\rangle|_{\mr{Sp}(n)}=2\langle \tilde u, \tilde
v\rangle|_{\mr{S}^{4n-1}}.
\end{eqnarray}

Let $\bar u$, $\bar v$ be the image of $u$, $v$ respectively under
the linearlization at $I$ of the quotient map
\begin{eqnarray} {\mr{Sp}}(n)
&\to& \bb HP^{n-1}\cr A&\mapsto&[A\tilde p]\;.\nonumber
\end{eqnarray}
One can see that $u$, $v$ are respectively the horizontal lift of
$\bar u$, $\bar v$, then, by definition,
$$ \fbox{$\langle
\bar u, \bar v\rangle_q=\langle u, v\rangle|_{\mr{Sp}(n)}$\;.}
$$
Here $ \langle \;, \rangle_q$ denote the quotient metric on ${\bb
HP}^{n-1}={\mr{Sp}(n)\over \mr{Sp}(n-1)\times \mr{Sp}(1)}$.

On the other hand, one can see that $\tilde u$ and $\tilde v$ are
respectively the horizontal lift of $\bar u$ and $\bar v$ in fiber
bundle $\mr{S }^{4n-1} \to \bb HP^{n-1}$, then, by definition,
$$ \fbox{$\langle \bar u, \bar v\rangle_{FS}=\langle \tilde u, \tilde
v\rangle|_{\mr{S}^{4n-1}}$\; .}
$$

Eq. (\ref{2wice}) then implies that $ \langle \bar u, \bar
v\rangle_q= 2\langle \bar u, \bar v\rangle_{FS}$. In view of the
fact that the Laplace operator gets multiplied by $1/a$ if the
Riemannian metric gets multiplied by a number $a$, Eq.
(\ref{LapFor}) follows from Lemma A.1 in Ref. \cite{meng08b}.

\end{proof}
We are now ready to introduce the notion of  $\mr{Sp}(1)$-Kepler
problems.

\begin{Def}\label{Def:1st}  Let $n\ge 2$ be an integer and
$\sigma$ an irreducible representation of $\mr{Sp}(1)$. The
$\mr{Sp}(1)$-Kepler problem in dimension $(4n-3)$ with magnetic
charge $\sigma$ is the quantum mechanical system for which the wave
functions are smooth sections of $\gamma_\sigma$, and the
hamiltonian is
\begin{eqnarray}\label{H:1st}
H=-{1\over 8\rho}\Delta_{\mathcal A}{1\over \rho}+{\bar \sigma(\bar
\sigma+2)+6(n-{7\over 8})\over 8\rho^4}-{1\over \rho^2}
\end{eqnarray}
where $\Delta_{\mathcal A}$ is the (non-positive) Laplace operator
on $\widetilde{\bb HP^n}$ twisted by $\gamma_\sigma$, $\bar\sigma$
is the highest weight of $\sigma$, and $\rho([Z])=|Z|$.
\end{Def}

Note that $\bb R^5_*$ and $\widetilde{\bb HP^2}$ are diffeomorphic.
We use $(r, \Theta)$ to denote the polar coordinates on $\bb R^5_*$
and $(\rho, \Phi)$ to denote the polar coordinates on
$\widetilde{\bb HP^2}$. Let $\pi$: $\widetilde{\bb HP^2}\to \bb
R^5_*$ be the diffeomorphism such that $\pi(\rho, \Phi)=(\rho^2,
\Phi)$, then
$$
\pi^*(dr^2+r^2\,d\Theta^2)=4\rho^2 (d\rho^2+ds^2_{FS})\quad
\mbox{and}\quad \pi^*(\mr{vol}_{{\bb
R^5}_*})=(2\rho)^5\mr{vol}_{\widetilde{\bb HP^2}}.
$$

Let $\gamma(\bar \sigma)$ be the pullback of $\gamma_\sigma$ by
$\pi^{-1}$. It is clear that $\gamma(\bar \sigma)$ is a hermitian
bundle over $\bb R^5_*$ with a natural hermitian connection $A$.
Recall from Ref. \cite{meng05} that \emph{the five dimensional
generalized MICZ-Kepler problem with magnetic charge $\mu \in
{1\over 2}\bb Z_{\ge 0}$} is the quantum mechanical system for which
the wave functions are smooth sections of $\gamma(\bar \sigma)$
($\mu=\bar\sigma/2$), and the hamiltonian is
$$\hat h_\mu=-{1\over 2}\Delta_A+{\mu^2+\mu\over 2r^2}-{1\over r}$$ where $\Delta_A$ is
the (non-positive) Laplace operator on  $\bb R^5_*$ twisted by
$\gamma(\bar \sigma)$ and $r(x)=|x|$. We are now ready to state the
following
\begin{Prop} The $\mr{Sp}(1)$-Kepler problem in dimension five with magnetic charge
$\sigma$ is equivalent to the five dimensional generalized
MICZ-Kepler problem with magnetic charge $\bar \sigma/2$.
\end{Prop}
\begin{proof}
Let $\Psi_i$ ($i=1$ or $2$) be a wave-section for the five
dimensional generalized MICZ-Kepler problem with magnetic charge
${\bar\sigma\over 2}$, and
$$\psi_i(\rho, \Phi):=(2\rho)^{5\over 2}\,\pi^*(\Psi_i)(\rho,
 \Phi)=(2\rho)^{5\over 2}\,\Psi_i(\rho^2, \Phi).$$ Then it is not
hard to see that
$$\displaystyle \int_{\widetilde{\bb
HP^2}}\overline{\psi_1}\psi_2\,\mr{vol}_{\widetilde{\bb
HP^2}}=\displaystyle \int_{\widetilde{\bb
HP^2}}\overline{\pi^*(\Psi_1)}\pi^*(\Psi_2)\, \pi^*(\mr{vol}_{\bb
R^5_*})=\displaystyle \int_{\bb R^5_*}\overline{\Psi_1}
\Psi_2\,\mr{vol}_{\bb R^5_*}$$ and
\begin{eqnarray}\displaystyle
\displaystyle\int_{\widetilde{\bb HP^2}}\overline{\psi_1}H\psi_2\,
\mr{vol}_{\widetilde{\bb HP^2}} &=&
\displaystyle\int_{\widetilde{\bb
HP^2}}\overline{\pi^*(\Psi_1)}\,{1\over \rho^{5\over
2}}H\rho^{5\over 2}\,\pi^*(\Psi_2) \,\pi^*(\mr{vol}_{\bb
R^5_*})\cr&=& \displaystyle\int_{\bb R^5_*}\overline{\Psi_1}\,\hat
h_{\bar\sigma}\,\Psi_2 \,\mr{vol}_{\bb R^5_*}. \nonumber
\end{eqnarray}Here we have used the fact that
\begin{eqnarray}
{1\over \rho^{5\over 2}}H\rho^{5\over 2} &=& -{1\over
8\rho^{7/2}}\left({1\over
\rho^4}\partial_\rho\rho^4\partial_\rho-{1\over \rho^2}\Delta_
{\mathcal A}|_{\bb HP^1}\right)\rho^{3/2}+{\bar \sigma(\bar
\sigma+2)+{27\over 4}\over 8\rho^4}-{1\over \rho^2}\cr &=& -{1\over
2}\left({1\over r^4}\partial_rr^4\partial_r-{1\over
r^2}\Delta_A|_{\mr S^4}\right)+{\bar \sigma(\bar \sigma+2)\over 8
r^2}-{1\over r}\cr &=&-{1\over 2}\Delta_A+{(\bar \sigma/2)^2+\bar
\sigma/2\over 2r^2}-{1\over r}=\hat h_{\bar\sigma/2}.\nonumber
\end{eqnarray}

\end{proof}
\begin{rmk}
In other words, we have proved that our $\mr{Sp}(1)$-Kepler problem
in dimension five is equivalent to (the quantum version of) the
Iwai's $\mr{SU}(2)$-Kepler problem \cite{Iwai90}.
\end{rmk}

\section{The dynamical symmetry analysis}\label{S:analysis}
Let $\psi$ be the eigenfunction of $H$ in Eq. (\ref{H:1st}) with
eigenvalue $E$, so $\psi$ is square integrable with respect to
volume form $\mr{vol}_{\widetilde{\bb HP^n}}$, and
\begin{eqnarray}\label{E:eigenvalue}
\left(-{1\over 8\rho}\Delta_{\mathcal A}{1\over \rho}+{\bar
\sigma(\bar \sigma+2)+6(n-{7\over 8})\over 8\rho^4}-{1\over
\rho^2}\right)\psi=E\psi.
\end{eqnarray}
We shall solve this eigenvalue problem by separating the angles from
the radius.

The branching rule for $({\mr {Sp}}(n), {\mr {Sp}}(n-1)\times \mr
{Sp}(1))$ plus the Frobenius reciprocity theorem imply that, as
module of $\mr {Sp}(n)$,
\begin{eqnarray}\label{Decomposition}
L^2(\gamma_\sigma|_{{\bb H}P^{n-1}})=\hat\bigoplus_{l=0}^\infty {\ms
R}_l(\sigma) \end{eqnarray} where ${\ms R}_l(\sigma)$ is the
irreducible and unitary representation of $\mr{Sp}(n)$ with the
highest weight $(l+\bar\sigma, l, 0,\cdots, 0)$.

Let $\{Y_{{l\bf m}}\mid {\bf m}\in {\mathcal I}(l)\}$ be a minimal
spanning set for ${\ms R}_l(\sigma)$. Write $\psi(x)=\tilde
R_{kl}(\rho)Y_{l\bf m}(\Phi)$. After separating out the angular
variables with the help of Eq. (\ref{LapFor}) in Lemma \ref{1stLem},
Eq. (\ref{E:eigenvalue}) becomes
\begin{eqnarray}
\left(-{1\over
8\rho^{4n-3}}\partial_\rho\rho^{4n-4}\partial_\rho{1\over
\rho}+{(l+{\bar\sigma\over 2})^2+(2n-1)(l+{\bar\sigma\over
2})+{3\over 2}(n-{7\over 8})\over 2\rho^4}-{1\over
\rho^2}\right)\tilde R_{kl}=E\tilde R_{kl}.\nonumber
\end{eqnarray} where $\tilde R_{kl}\in L^2({\bb R}_+, \rho^{4n-4}\,d\rho)$.
Let $R_{k(l+{\bar\sigma\over 2})}(t)=\tilde R_{kl}(\sqrt
t)/t^{5/4}$, then we have $ R_{k(l+{\bar\sigma\over 2})}\in L^2({\bb
R}_+, t^{2n}\,dt)$ and
\begin{eqnarray}
\left( -{1\over 2t^{2n}}\partial_t
t^{2n}\partial_t+{(l+{\bar\sigma\over
2})^2+(2n-1)(l+{\bar\sigma\over 2})\over 2t^2}-{1\over t}\right)
R_{k(l+{\bar\sigma\over 2})}=E R_{k(l+{\bar\sigma\over
2})}.\nonumber
\end{eqnarray}
By quoting results from appendix A in Ref. \cite{meng08a}, we have
\begin{eqnarray}
E_{k(l+{\bar\sigma\over 2})}=-{1/2\over (k+l+{\bar\sigma+2n\over
2}-1)^2}
\end{eqnarray} where $k=1, 2, 3, \cdots$.
Let $I=k-1+l$, then the {\em bound energy spectrum} is
\begin{eqnarray}
E_I=-{1/2\over (I+n+{\bar\sigma\over 2})^2} \end{eqnarray} where
$I=0, 1, 2, \cdots$. This proves part 1) of the main theorem.
Moreover, since $\tilde R_{kl}(\rho)=\rho^{5\over
2}R_{k(l+{\bar\sigma\over 2})}(\rho^2)$, we have
$$
\tilde R_{kl}(\rho)=c(k,l+{\bar\sigma\over
2})\rho^{2l+\bar\sigma+5/2}L^{2l+\bar\sigma+2n-1}_{k-1}\left({2\rho^2\over
I+n+{\bar\sigma\over 2}}\right)\exp\left(-{\rho^2\over
I+n+{\bar\sigma\over 2}}\right).
$$

For each integer $I\ge 0$, we let ${\ms H}_I(\sigma)$ be the linear
span of
$$
\{\tilde R_{kl}Y_{l{\bf m}}\mid {\bf m}\in {\mathcal I}(l),
k-1+l=I\},
$$ then
\begin{eqnarray}\label{energyE} {\ms H}_I(\sigma)\cong \bigoplus_{l=0}^{I} {\ms
R}_l(\sigma) \end{eqnarray} is the eigenspace of $H$ with eigenvalue
$E_I$, and the Hilbert space of bound states admits the following
orthogonal decomposition into the eigenspaces of $H$:
$$
{\ms H}(\sigma)=\hat\bigoplus_{I=0}^\infty {\ms H}_I(\sigma).
$$
Part 4) of the main theorem is then clear. We shall show that ${\ms
H}_I(\sigma)$ is the highest weight representation of $\tilde{\mr
U}(2n)$ with highest weight $$(\underbrace{-1, \cdots,  -1}_{2n-2},
-(1+I), -(1+I+\bar\sigma)).$$  To do that, we need to twist the
Hilbert space of bound states and the energy eigenspaces.
\subsection{Twisting}
Let $n_I=I+n+{|\bar\sigma|\over 2}$ for each integer $I\ge 0$. For
each $\psi_I\in {\ms H}_I$, as in Refs. \cite{Barut71,MZ07}, we
define its twist $\tilde \psi_I$ by the following formula:
\begin{eqnarray}
\tilde \psi_I(Z)=c_I{1\over |Z|^{5\over 2}}\psi_I(\sqrt{n_I\over
2}[Z])
\end{eqnarray} where $c_I$ is the unique constant such that
$$\int_{{\bb
H^n_*}}|\tilde \psi_I|^2=\int_{\widetilde{\bb HP^n}}|\psi_I|^2.$$

We use $\tilde {\ms H}_I(\sigma)$ to denote the span of all such
$\tilde\psi_I$'s, $\tilde {\ms H}(\sigma)$ to denote the Hilbert
space direct sum of $\tilde{\ms H_I}(\sigma)$. We write the linear
map sending $\psi_I$ to $\tilde\psi_I$ as
\begin{eqnarray}\label{taumap}
\tau:\quad {\ms H}(\sigma)\to \tilde{\ms H}(\sigma).
\end{eqnarray}
Then $\tau$ is a linear isometry.

Since
$$
H\psi_I=E_I\psi_I,
$$
after re-scaling: $[Z]\to \sqrt{n_I\over 2}[Z]$, we have
\begin{eqnarray}
&&\left(-{(2/n_I)^2\over 8r}\Delta_{\mathcal A}{1\over
r}+(2/n_I)^2{\bar \sigma(\bar\sigma+2)+6(n-{7\over 8})\over
8r^4}-{2/n_I\over r^2}\right)\psi_I(\sqrt{n_I\over 2}[Z])\cr
&=&E_I\psi_I(\sqrt{n_I\over 2}[Z]),\nonumber \end{eqnarray} where
$r=|Z|$. Multiplying by $(n_Ir)^2$, we obtain
\begin{eqnarray}
{1\over r^{5/2}}\left(-{r\over 2}\left(\Delta_{\mathcal A}-{\bar
\sigma(\bar \sigma+2)+6(n-{7\over 8})\over r^2}\right){1\over
r}-{2n_I}\right)r^{5/2}\tilde\psi_I(Z)&= &
n_I^2E_Ir^2\tilde\psi_I(Z)\cr &=&-{1\over 2}
r^2\tilde\psi_I(Z).\nonumber
\end{eqnarray} Applying Lemma A.1 in Ref. \cite{meng08b}, the previous equation becomes
\begin{eqnarray}
-{1\over 2} r^2\tilde\psi_I&=&\left(-{1\over 2}\left({1\over
r^{4n-{5\over 2}}}\partial_r r^{4n-4}\partial_r r^{3\over
2}-{\Delta\mid _{{\mr S}^{4n-1}}+6(n-{7\over 8})\over
r^2}\right)-{2n_I}\right)\tilde\psi_I\cr &=& \left(-{1\over
2}\left({1\over r^{4n-1}}\partial_r r^{4n-1}\partial_r-{\Delta\mid
_{{\mr S}^{4n-1}}\over r^2}\right)-{2n_I}\right)\tilde\psi_I\cr &=&
\left(-{1\over 2}\Delta-2n_I\right)\tilde \psi_I, \nonumber
\end{eqnarray} where $\Delta$ is the (negative definite) standard Laplace operator on $\bb
H^n$, and $\tilde \psi_I$ is viewed as a $\sigma$-valued function on
$\bb H^n_*$. Then
\begin{eqnarray}\label{Osc}\left(-{1\over 2}\Delta+{1\over 2} r^2\right)\tilde
\psi_I=(2I+|\bar\sigma|+2n)\tilde \psi_I.
\end{eqnarray}

Denote the collection of all smooth sections of $\gamma_\sigma$ by
$C^\infty(\gamma_\sigma)$, and the collection of all smooth
complex-valued functions on $\bb H^n_*$ by $C^\infty(\bb H^n_*)$.
Note that an element in $\sigma^*\otimes C^\infty(\gamma_\sigma)$
can be viewed as a smooth map $F$: $\bb H^n_*\to \mr{End} (\sigma)$
such that $F(x\cdot g^{-1})=\rho_\sigma(g)\circ F(x)$ for any $g\in
\mr{Sp}(1)$. Denote by $\mr{Tr}\,F$ the map sending $x\in \bb H^n$
to the trace of $F(x)$ and by $\mr{Tr}$ the linear map sending
$F\in\sigma^*\otimes C^\infty(\gamma_\sigma)$ to $\mr{Tr}\,F\in
C^\infty(\bb H^n_*)$.

The left $\mr{Sp}(n)$-action on $\bb H^n_*$ induces natural actions
on both $C^\infty(\gamma_\sigma)$ (hence on $\sigma^*\otimes
C^\infty(\gamma_\sigma)$) and $C^\infty(\bb H^n_*)$ with respect to
which $\mr{Tr}$ is equivariant: For any $g\in\mr{Sp}(n)$,
$$
\mr{Tr}\left(L_g^*F\right)(x)=\mr{Tr}\left(L_g^*F(x)\right)=\mr{Tr}\left(F(g\cdot
x )\right)=\mr{Tr}\,F(g\cdot x)=L_g^*\left(\mr{Tr}\,F\right)(x).
$$

The right $\mr{Sp}(1)$-action on $\bb H^n_*$ induces a natural
action on $C^\infty(\bb H^n_*)$. The action of $\mr{Sp}(1)$ on
$\sigma^*$ yields a natural $\mr{Sp}(1)$-action on $\sigma^*\otimes
C^\infty(\gamma_\sigma)$ for which
$$
(g\cdot F) (x)=F(x)\circ \rho_\sigma(g^{-1})\quad\mbox{for any
$g\in\mr{Sp}(1)$}.
$$
It is clear that $\mr{Tr}$ is also $\mr{Sp}(1)$-equivariant: For any
$g\in\mr{Sp}(1)$,
\begin{eqnarray}
\mr{Tr}\left(g\cdot F\right)(x)&=& \mr{Tr}\left((g\cdot F)(x)\right)
\cr &= & \mr{Tr}\left(F(x)\circ\rho_\sigma(g^{-1}) \right)\cr &=
&\mr{Tr}\left(\rho_\sigma(g^{-1})\circ F(x) \right)\cr &=
&\mr{Tr}\left(F(x \cdot g)\right)=\mr{Tr}\,F(x \cdot g)\cr
&=&R_g^*(\mr{Tr}\,F)(x)\nonumber
\end{eqnarray} where $R_g$ stands for the right action of $g$
on $\bb H^n_*$.

The inner products on $\sigma^*$ and $C^\infty(\gamma_\sigma)$
yields the tensor product inner product on $\sigma^*\otimes
C^\infty(\gamma_\sigma)$:
$$
\langle\alpha\otimes \psi, \beta\otimes \phi\rangle=\langle\alpha,
\beta \rangle\int_{\bb H^n_*} \langle \psi, \phi\rangle=\int_{\bb
H^n_*} \mr{Tr}\,\left((\alpha\otimes \psi)^\dag (\beta\otimes
\phi)\right).
$$
On $C^\infty(\bb H^n_*)$, there is a natural inner product:
$$
\langle f, g\rangle=\int_{\bb H^n_*} \bar f g.
$$
\begin{Lem} Let $\chi_\sigma$ be the character function of $\sigma$ on $\mr{Sp}(1)$,
$$||\chi_\sigma||^2={\int_{\mr{Sp}(1)}|\chi_\sigma(g)|^2\, dg\over \int_{\mr{Sp}(1)} dg}.$$
Then $$||\chi_\sigma||^2\langle F, G\rangle =\dim \sigma\,\langle
\mr{Tr}\, F, \mr{Tr}\, G\rangle.
$$Consequently $\mr{Tr}$ is injective.
\end{Lem}
\begin{proof} Since
\begin{eqnarray}
\langle \mr{Tr}\, F, \mr{Tr}\, G\rangle &= & \int_{\bb H^n_*}
\overline{\mr{Tr}\, F}\, \mr{Tr}\, G=\int_{\bb H^n_*} \mr{Tr}\,
F^\dag\, \mr{Tr}\, G,\nonumber\\
\langle F, G\rangle &=&\int_{\bb H^n_*} \mr{Tr}\,\left(F^\dag
G\right),\nonumber
\end{eqnarray} we just need to show that
\begin{eqnarray}
||\chi_\sigma||^2\int_{\bb H^n_*} \mr{Tr}\,\left(F^\dag
G\right)=\dim\sigma\int_{\bb H^n_*} \mr{Tr}\, F^\dag\, \mr{Tr}\, G
\end{eqnarray} for any $F$, $G$ in $\sigma^*\otimes
C^\infty(\gamma_\sigma)$. The two integrants are definite not equal,
however, their integrations over $\bb H^n_*$ are equal, that is
because their integrations along any orbit of the right action of
$\mr{Sp}(1)$ on $\bb H^n_*$ are equal:
\begin{Claim} For any $F$, $G$ in $\sigma^*\otimes
C^\infty(\gamma_\sigma)$, and $Z_0\in \bb H^n_*$, we have
\begin{eqnarray}
||\chi_\sigma||^2\int_{\mr{Sp}(1)} \mr{Tr}\,\left(F^\dag(Z_0\cdot g)
G(Z_0\cdot g)\right)\,dg=\dim\sigma\int_{\mr{Sp}(1)} \mr{Tr}\,
F^\dag(Z_0\cdot g)\, \mr{Tr}\, G(Z_0\cdot g)\, dg.\nonumber
\end{eqnarray}
\end{Claim}This claim follows easily from Lemma \ref{simplefact}.
Here the fact that $dg$ is bi-invariant must be used in the proof.
\end{proof}

This lemma implies that $\mr{Tr}$ is injective. Actually, we have
\begin{Prop} $\mr{Tr}$ induces a Hilbert space isomorphism
\begin{eqnarray}\label{corres}
 \mr{Tr}_*:\quad\hat \bigoplus_{\sigma^* \in
\widehat {\mr {Sp}(1)}}\sigma^*\otimes\tilde{\ms H}(\sigma)
&\cong&L^2\left(\bb H^n_*\right) =L^2\left(\bb H^n\right)\\
\oplus_\sigma F_\sigma & \mapsto
&\oplus_\sigma{\sqrt{\dim\sigma}\over ||\chi_\sigma||}\mr{Tr}\,
F_\sigma\nonumber
\end{eqnarray} which is equivariant with respect to both the
$\mr{Sp}(n)$-actions and the $\mr{Sp}(1)$-actions.
\end{Prop}
\begin{proof}
Let ${\mathcal H}_k $ be the $k$-th energy eigenspace of the
$4n$-dimensional isotropic harmonic oscillator with hamiltonian
$-{1\over 2}\Delta+{1\over 2} r^2$. In view of Eq. (\ref{Osc}), we
have the following map
\begin{eqnarray}\label{claimEq}
\iota:=\oplus\mr{Tr}:\quad
\bigoplus_{2I+\bar\sigma=k}\sigma^*\otimes\tilde{\ms
H}_I(\sigma)\longrightarrow {\mathcal H}_k.\end{eqnarray}

\begin{Claim} $\iota$ is an isomorphism.
\end{Claim}
\begin{proof} Since $\iota$ is injective, one just needs to verify the {\em
dimension equality}:
$$
\sum_{2I+\bar\sigma=k} \dim\sigma^* \cdot \dim {\ms H}_I(\sigma)=
\dim {\mathcal H}_k.
$$ In view of Eq. (\ref{energyE}),
\begin{eqnarray}
\dim \tilde{\ms H}_I(\sigma)=\dim {\ms H}_I(\sigma)=\sum_{l=0}^{I}
\dim {\ms R}_l(\sigma).\nonumber
\end{eqnarray} Using the dimension formula in representation theory to compute $\dim {\ms
R}_l(\sigma)$, one arrives at the following explicit form of the
dimension equality:
\begin{eqnarray} {\sum_{q\le I,\, q\le
p}^{2I+p-q=k}}{(1+p-q)^2\over 1+p}\left(1+{p+q\over
2n-1}\right)\left(\begin{matrix}p+2n-2\cr
p\end{matrix}\right)\left(\begin{matrix}q+2n-3\cr
q\end{matrix}\right)=\left(\begin{matrix}4n+k-1\cr
4n-1\end{matrix}\right).\nonumber
\end{eqnarray}
Note that the left hand side of this dimension equality can be
simplified into the following more attractive form:
$$
{\sum_{0\le q\le I,\, q<
p}^{2I-1+p-q=k}}(p-q)\left(\begin{matrix}p+2n-2\cr
2n-1\end{matrix}\right)\left(\begin{matrix}q+2n-3\cr
2n-3\end{matrix}\right)+<p\leftrightarrow q>,
$$ so its generating function, when multiplied by $(1-t^2)$, is equal
to
\begin{eqnarray}
& & \sum_{q< p}t^{p+q-1}
 \left[(p-q)\left(\begin{matrix}p+2n-2\cr
2n-1\end{matrix}\right)\left(\begin{matrix}q+2n-3\cr
2n-3\end{matrix}\right)+<p\leftrightarrow q>\right]\cr &=& {1\over
2}\sum_{p, q\ge 0}t^{p+q-1}\left[(p-q)\left(\begin{matrix}p+2n-2\cr
2n-1\end{matrix}\right)\left(\begin{matrix}q+2n-3\cr
2n-3\end{matrix}\right)+<p\leftrightarrow q>\right]\cr &=& \sum_{p,
q\ge 0}t^{p+q-1}\left[(p-q)\left(\begin{matrix}p+2n-2\cr
2n-1\end{matrix}\right)\left(\begin{matrix}q+2n-3\cr
2n-3\end{matrix}\right)\right]\cr &=& \sum_{p, q\ge
0}t^{p+q-1}\left[p\left(\begin{matrix}p+2n-2\cr
2n-1\end{matrix}\right)\left(\begin{matrix}q+2n-3\cr
2n-3\end{matrix}\right)-q\left(\begin{matrix}p+2n-2\cr
2n-1\end{matrix}\right)\left(\begin{matrix}q+2n-3\cr
2n-3\end{matrix}\right)\right]\cr &=&(1-t)^{2-2n}\sum_{p\ge
0}pt^{p-1}\left(\begin{matrix}p+2n-2\cr
2n-1\end{matrix}\right)-(1-t)^{-2n}\sum_{q\ge
0}qt^q\left(\begin{matrix}q+2n-3\cr 2n-3\end{matrix}\right) \cr
&=&(1-t)^{2-2n}\left(t(1-t)^{-2n}\right)'-(1-t)^{-2n}t\left((1-t)^{2-2n}\right)'\cr
&=&(1-t)^{-4n}(1-t^2).\nonumber
\end{eqnarray}Therefore, the generating function for the left hand side of dimension equality is
$(1-t)^{-4n}$. Since the generating function for the right hand side
of dimension equality is obviously $(1-t)^{-4n}$, we have proved the
dimension equality, hence the claim.
\end{proof}
As a consequence of the above claim, we have the following Hilbert
space isomorphism:
\begin{eqnarray}
\hat \bigoplus_{\sigma^* \in \widehat {\mr
{Sp}(1)}}\sigma^*\otimes\tilde{\ms H}(\sigma)&= &\hat
\bigoplus_{\sigma \in \widehat {\mr{Sp}(1)}, I\ge 0}
\sigma^*\otimes\tilde{\ms H}_I(\sigma)\cr
&\buildrel\mr{Tr}_*\over\cong&\hat \bigoplus_{k=0}^\infty {\mathcal
H}_k\cr &=&L^2\left(\bb H^n_*\right).\nonumber
\end{eqnarray}
\end{proof}
\subsection{Proof of Theorem \ref{T:main} }
Parts 1) and 4) have been proved. Let $(q^i, p_i)$'s be the
canonical symplectic coordinates on $T^*\bb H^n$, in terms of which,
the canonical symplectic form can be written as
$$\omega=dq^i\wedge dp_i\,.$$
The complex structures $I$, $J$ and $K$ on $\bb H^n$ are
diffeomorphisms from $\bb H^n$ to itself, so $T^*I$, $T^*J$ and
$T^*K$ are symplectomorphisms from $(T^*\bb H^n, \omega)$ to itself.
Introducing $z^i=q^{4i-3}+\vec iq^{4i-2}+\vec j q^{4i-1}+\vec k
q^{4i}$ and $w_i=p_{4i-3}+\vec ip_{4i-2}+\vec j p_{4i-1}+\vec k
p_{4i}$, one can check that $$T^*I(z^i, w_i)=\vec i(z^i, w_i),\quad
T^*J(z^i, w_i)=\vec j(z^i, w_i), \quad T^*K(z^i, w_i)=\vec k(z^i,
w_i),$$ so $(T^*I, T^*J, T^*K)$ is a quaternionic structure on
$T^*\bb H^n$ and $(z^1,\cdots, z^n, w_1,\cdots, w_n)$ are the
standard quaternionic coordinates on $T^*\bb H^n$. Let $z=(z^1,
\cdots, z^n)^T$ and $w=(w_1, \cdots, w_n)^T$, then
\begin{eqnarray}\omega = \mr{Re}\left( dz^\dag\wedge d
w\right). \end{eqnarray}

Let $q\in \mr{Sp}(1)$. The right action of $q\in \mr{Sp}(1)$ on $\bb
H^n$ induces a right action on $T^*\bb H^n$ which maps $(z^T,w^T)$
to $(z^Tq, w^Tq)=(z^1q, \cdots, z^nq, w_1q, \cdots, w_nq)$, so it
leaves $\omega$ invariant because
$$\mr{Re}\left( q^{-1}(d\bar z^k\wedge d w_k)q\right)=\mr{Re}\left(
d\bar z^k\wedge d w_k\right).$$
Let
$$
G=\left\{g\in \mr{End}_{\bb R}(T^*\bb H^n)\mid g^*\omega=\omega,
\mbox{$g$ commutes with the right $\mr{Sp}(1)$ action} \right\}.
$$
We \underline{claim} that $G=\mr{O}^*(4n)$. To prove that, we assume
that $\vec i\,\vec j=-\vec k$, and then rewrite the quaternion
vectors $z=z'+\vec jz''$ and $w=w'+\vec jw''$ as complex vectors
$$
Z=\left(\begin{matrix}z' \cr  z''\end{matrix}\right),\quad
W=\left(\begin{matrix}w' \cr  w''\end{matrix}\right)
$$ respectively.
The right action by $\bb H$ can be written as follows:
\begin{eqnarray}
Z\cdot \vec i=iZ,\; W\cdot \vec i=iW,\; Z\cdot \vec j=J_{2n}\bar
Z,\; W\cdot \vec j=J_{2n}\bar W,\nonumber
\end{eqnarray} where
$$
J_{2n}=\left(\begin{matrix}0 & -I_n\cr I_n & 0\end{matrix}\right)
$$ with $I_{2n}$ being the identity matrix of order $2n$. Let
$$U={Z+iW\over \sqrt 2}, \quad V={Z-iW\over \sqrt 2},$$
then the standard symplectic form can be written as
\begin{eqnarray}
\omega &=& {1\over 2}(dZ^\dag, dW^\dag)\left(\begin{matrix}0 &
I_{2n}\cr -I_{2n} &
0\end{matrix}\right)\wedge\left(\begin{matrix}dZ\cr
dW\end{matrix}\right)\cr &=& {1\over 2i}(d U^\dag, d
V^\dag)\wedge\left(\begin{matrix}I_{2n} & 0\cr 0 &
-I_{2n}\end{matrix}\right)\left(\begin{matrix}dU\cr
dV\end{matrix}\right);\nonumber
\end{eqnarray}
moreover,
\begin{eqnarray}
U\cdot \vec i=iU,\; V\cdot \vec i=iV,\; U\cdot \vec j=J_{2n}\bar
V,\; V\cdot \vec j=J_{2n}\bar U.\nonumber
\end{eqnarray}

Therefore, $g\in G$ if and only if $g\in \mr{End}_{\bb R}(T^*\bb
H^n)$ is complex linear and
\begin{eqnarray}
g^\dag \left(\begin{matrix}I_{2n}& 0\cr 0 & -I_{2n}
\end{matrix}\right) g= \left(\begin{matrix} I_{2n} & 0\cr 0 & -I_{2n}
\end{matrix}\right),\quad g \left(\begin{matrix} 0 & J_{2n}\cr J_{2n}
& 0
\end{matrix}\right)=\left(\begin{matrix} 0 & J_{2n}\cr J_{2n}&
0 \end{matrix}\right) \bar g,\nonumber
\end{eqnarray}
or equivalently
\begin{eqnarray}
g^\dag \left(\begin{matrix}I_{2n}& 0\cr 0 & -I_{2n}
\end{matrix}\right) g= \left(\begin{matrix} I_{2n} & 0\cr 0 & -I_{2n}
\end{matrix}\right),\quad g^T \left(\begin{matrix}0 & J_{2n}\cr
-J_{2n} & 0\end{matrix}\right)g=\left(\begin{matrix} 0 & J_{2n}\cr
-J_{2n} & 0 \end{matrix}\right).\nonumber
\end{eqnarray}
So we conclude that $G=\mr{U}(2n, 2n)\cap\mr{O}(4n, \bb C)$, i.e.,
$G=\mr{O}^*(4n)$.

It is now clear that Eq. (\ref{corres}) is really the decomposition
in Eqs. (4.1) and (4.2) in Ref. \cite{Howe} with the dual pair being
$(\mr{Sp}(1), \mr{O}^*(4n))$ inside $\mr{Sp}_{8n}(\bb R)$; as a
consequence, $\tilde{\ms H}(\sigma)$ is a unitary highest weight
module of $\widetilde{\mr {O}^*}(4n)$. In view of the fact that
$\tau$ in Eq. (\ref{taumap}) is an isometry, by pulling back the
action of $\widetilde{\mr {O}^*}(4n)$ on $\tilde {\ms H}(\sigma)$
via $\tau$, we get the action of $\widetilde{\mr {O}^*}(4n)$ on
${\ms H}(\sigma)$. Part 5) is then proved.

\vskip 5pt
Since
\begin{eqnarray}
\omega = {1\over 2i}(d U^\dag, d V^T)\left(\begin{matrix}I_{2n} &
0\cr 0 & I_{2n}\end{matrix}\right)\wedge\left(\begin{matrix}dU\cr
d\bar V\end{matrix}\right),
\end{eqnarray} we know that  $\omega$ is invariant under the action
of $\mr{U}(4n)$ on $T^*\bb H^n=\bb C^{2n}\oplus \overline{\bb
C^{2n}}$. Note that this $\mr{U}(4n)$, being the unitary group which
leaves
$$|U|^2+|\bar V|^2=|Z|^2+|W|^2=|z|^2+|w|^2$$ invariant, is a maximal
compact subgroup of $\mr{Sp}_{8n}(\bb R)$. It is not hard to see that
\begin{eqnarray}\label{maxcpt}\mr{U}(4n)\cap
\mr{O}^*(4n)=\mr{U}(2n)\supseteq \mr{Sp}(n)
\end{eqnarray} with the following identification:
\begin{eqnarray}\label{reverse}
\mr{U}(4n)\ni\left(\begin{matrix}A & 0\cr 0 &
-J_{2n}AJ_{2n}\end{matrix}\right)\longleftrightarrow
\left(\begin{matrix}A & 0\cr 0 & -J_{2n}\bar A
J_{2n}\end{matrix}\right)\in \mr{O}^*(4n).
\end{eqnarray}
A consequence of Eq. (\ref{reverse}) is that the imbedding ${\mr
U}(2n)\subset {\mr U}(4n)$ maps
$$
\mr{diag}\{\underbrace{0, \cdots, 0}_{i-1}, 1, 0,\cdots, 0\}\in
\frk{u}(2n)
$$
into
$$
\mr{diag}\{\underbrace{0, \cdots, 0}_{i-1}, 1, 0,\cdots,
0\}+\mr{diag}\{\underbrace{0, \cdots, 0}_{\bar i-1}, 1, 0,\cdots,
0\}\in \frk{u}(4n)
$$  where $\bar i>2n$ is the unique index such that $|\bar i-
i-2n|=n$.

Note that $\mr{Sp}(n)$ acts on $\bb H^n$ from left, and the induced
action on $T^*\bb H^n$ which maps $(z^T,w^T)$ to $(gz^T, gw^T)$ can
be identified with the subgroup of $\mr{U}(2n)$ (a maximal compact
subgroup of $\mr{O}^*(4n)$) which preserves the quaternionic
structure on $T^*\bb H^n$. The above unitary action of $\widetilde
{\mr O}^*(4n)$ on the Hilbert space ${\ms H}(\sigma)$ of bound
states, extends the manifest unitary action of $\mr{Sp}(n)$.

Since ${\mathcal H}_k$ is invariant under the action of $\tilde {\mr
U}(4n)$, in view of Eqs. (\ref{maxcpt}) and (\ref{corres}), the fact
that the Hilbert space isomorphism
\begin{eqnarray}\label{identify}
\mr{Tr}_*:\quad \bigoplus_{2I+\bar\sigma=k}\sigma^*\otimes\tilde{\ms
H}_I(\sigma)\to {\mathcal H}_k
\end{eqnarray} is an isomorphism of
$\tilde{\mr U}(2n)$-modules implies that $\tilde{\ms H}_I(\sigma)$
(hence ${\ms H}_I(\sigma)$) is a $\tilde{\mr U}(2n)$-module. Since
Eq. (\ref{energyE}) is a decomposition of $\mr{Sp}(n)$-modules, by
using the Littlewood branch rule for $(\mr{U}(2n), \mr{Sp}(n))$,
there is a half integer $\kappa$ such that
\begin{eqnarray}{\ms H}_I(\sigma)\cong  {\mathcal
R}_I^\kappa(\sigma) \; \mbox{or}\; \overline{{\mathcal
R}_I^\kappa(\sigma)}
\end{eqnarray} as irreducible representations of $\tilde{\mr
U}(2n)$. In view of Eq. (\ref{reverse}) and its succeeding
paragraph, the isomorphism in Eq. (\ref{identify}) and some simple
facts on harmonic isolators imply that ${\ms H}_I(\sigma)\cong
\overline{{\mathcal R}_I^\kappa(\sigma)}$ and
$$
2I+\bar \sigma +2n\kappa= 2I+\bar \sigma +2n,
$$
so we must have $\kappa=1$. This proves part 3), and consequently
part 2).

\appendix
\section{A Simple Fact}
Let $G$ be a compact simple group, $\sigma$ be an irreducible and
unitary representation of $G$. We also use $\sigma$ to denote the
underlying (complex) representation space of $\sigma$ and
$\mr{End}(\sigma)$ to denote the space of endomorphisms of $\sigma$.
\begin{Lem}\label{simplefact} Let $F$ be an Hermitian inner product on $\mr{End}(\sigma)$.
Suppose that $F$ is bi-invariant under the action of $G$:
$$
F(\rho_\sigma(g)X, \rho_\sigma(g)Y)=F(X, Y), \quad
F(X\rho_\sigma(g), Y\rho_\sigma(g))=F(X, Y)
$$ for any $g\in G$. Then
$$
F(X, Y)={\mr{Tr}(I, I)\over \dim\sigma}\mr{Tr}(X^\dag Y).
$$
\end{Lem}
This lemma is equivalent to the following statement in linear
algebra: \emph{ Let $V$ be an Hermitian vector space. Suppose that
$Q$ is an Hermitian inner product on $\mr{End}(V)$ such that $$Q(AX,
Y)=Q(X, AY), \quad Q(XA, Y)=Q(X, YA)$$ for any Hermitian operator
$A$ on $V$, then
$$
Q(X, Y)=\lambda\mr{Tr}(X^\dag Y)
$$ for some constant $\lambda$}. To prove it, we note that
$Q(AX, Y)=Q(X, A^\dag Y)$ and $Q(XA, Y)=Q(X,YA^\dag )$ for any $A\in
\mr{End}(V)$. Therefore, 1) $Q(X, Y)=Q(I,X^\dag Y)$, 2) $Q(I, Y)$ is
a trace: it is linear in $Y$ and
$$
Q(I, XY)=Q(X^\dag, Y)=Q(I, YX).
$$ Consequently, we have the statement proved.

\end{document}